\ifx\pdfoutput\undefined        % not running pdflatex
  \documentclass{iaus}
\else                           % running pdflatex
  \documentclass[pdftex]{iaus}
  \usepackage[bookmarks=false]{hyperref}
\fi

\usepackage{graphicx}
\usepackage{url}
  \checkfont{eurm10}
  \iffontfound
    \IfFileExists{upmath.sty}
      {\typeout{^^JFound AMS Euler Roman fonts on the system,
                   using the 'upmath' package.^^J}%
       \usepackage{upmath}}
      {\typeout{^^JFound AMS Euler Roman fonts on the system, but you
                   dont seem to have the}%
       \typeout{'upmath' package installed. iaus.cls can take advantage
                 of these fonts,^^Jif you use 'upmath' package.^^J}%
      }
  \else
  \fi

  \checkfont{msam10}
  \iffontfound
    \IfFileExists{amssymb.sty}
      {\typeout{^^JFound AMS Symbol fonts on the system, using the
                'amssymb' package.^^J}%
       \usepackage{amssymb}%
         \let\leq=\leqslant
         
      }{}
  \fi

  \IfFileExists{amsbsy.sty}
    {\typeout{^^JFound the 'amsbsy' package on the system, using it.^^J}%
     \usepackage{amsbsy}}
    {}

\newsavebox{\astrutbox}
\sbox{\astrutbox}{\rule[-5pt]{0pt}{20pt}}

\newcommand{\EQ}{\begin{equation}}
\newcommand{\EN}{\end{equation}}
\newcommand{\EQA}{\begin{eqnarray}}
\newcommand{\ENA}{\end{eqnarray}}

\newcommand{\Eq}[1]{eq.~(\ref{#1})}

\newcommand{\Sec}[1]{\S\,\ref{#1}}

\newcommand{\Fig}[1]{figure~\ref{#1}}

\newcommand{\bra}[1]{\langle #1\rangle}

\newcommand{\meanFF}{\overline{\mbox{\boldmath ${\cal F}$}} {}}

\newcommand{\meanF}{\overline{\cal F}}
\newcommand{\meanJ}{\overline{J}}
\newcommand{\meanU}{\overline{U}}

\newcommand{\meanBB}{\overline{\mbox{\boldmath $B$}}}
\newcommand{\meanUU}{\overline{\mbox{\boldmath $U$}}}
\newcommand{\meanJJ}{\overline{\mbox{\boldmath $J$}}}

\newcommand{\bb}{\mbox{\boldmath $b$} {}}

\newcommand{\BB}{\mbox{\boldmath $B$} {}}

\newcommand{\aaaa}{\mbox{\boldmath $a$} {}}
\newcommand{\jj}{\mbox{\boldmath $j$} {}}

\newcommand{\emf}{\mbox{\boldmath ${\cal E}$} {}}

\newcommand{\yan}[5]{~ #1~ #5. {\em Astron. Nachr. }{\bf #2}, #3--#4.}
\newcommand{\yana}[5]{~ #1~ #5. {\em Astron. Astrophys. }{\bf #2}, #3--#4.}

\newcommand{\ysph}[5]{~ #1~ #5. {\em Solar Phys. }{\bf #2}, #3--#4.}

\newcommand{\yjfm}[5]{~ #1~ #5. {\em J. Fluid Mech. }{\bf #2}, #3--#4.}

\newcommand{\yjgr}[5]{~ #1~ #5. {\em J. Geophys. Res. }{\bf #2}, #3--#4.}

\newcommand{\yapj}[5]{~ #1~ #5. {\em Astrophys. J. }{\bf #2}, #3--#4.}

\newcommand{\ybook}[3]{~ #1~ {\em #2}. #3.}

\newcommand{\sjour}[4]{~ #1~ #3. {\em #2} (submitted, #4).}

\newcommand{\pana}[3]{ ~#1~ ``#2,'' {\em Astron. Astrophys.} (in press, #3).}

\newcommand{\G}{\,{\rm G}}
\newcommand{\Gtwopers}{\,{\rm G^2\!/s}}

\newcommand{\s}{\,{\rm s}}
\newcommand{\mpers}{\,{\rm m/s}}

\newcommand{\cm}{\,{\rm cm}}

\newcommand{\km}{\,{\rm km}}

\newcommand{\Mx}{\,{\rm Mx}}

\title[Helical coronal ejections and their role in the solar cycle]
{Helical coronal ejections and their role in the solar cycle}

\author[A. Brandenburg et al.]%
{Axel Brandenburg$^1$, Christer Sandin$^2$ \and Petri J.\ K\"apyl\"a$^{3,4}$}

\affiliation{$^1$Nordita, Blegdamsvej 17, D-2100 Copenhagen \O, Denmark
\\[\affilskip]
$^2$Stockholm, Sweden
\\[\affilskip]
$^3$Kiepenheuer-Institut f\"ur Sonnenphysik, Sch\"oneckstra\ss{}e 6,
D-79104 Freiburg, Germany
\\[\affilskip]
$^4$Department of Physical Sciences, Astronomy Division, P.O. Box 3000,
FIN-90014 University of Oulu, Finland
}

\pubyear{2004}
\volume{223}
\pagerange{1--8}
\date{?? and in revised form ??}
\jname{Multi-Wavelength Investigations of Solar Activity}
\editors{A.V. Stepanov, E.E. Benevolenskaya \& A.G. Kosovichev, eds.}

\begin{document}
\maketitle

\begin{abstract}
The standard theory of the solar cycle in terms of an alpha-Omega dynamo
hinges on a proper understanding of the nonlinear alpha effect.
Boundary conditions play a surprisingly important role in determining
the magnitude of alpha.
For closed boundaries, the total magnetic helicity is conserved, and
since the alpha effect produces magnetic helicity of one sign in the
large scale field, it must simultaneously produce magnetic helicity of
the opposite sign.
It is this secondary magnetic helicity that suppresses the dynamo
in a potentially catastrophic fashion.
Open boundaries allow magnetic helicity to be lost.
Simulations are presented that allow an estimate of alpha in the
presence of open or closed boundaries, either with or without solar-like
differential rotation.
In all cases the sign of the magnetic helicity agrees with that observed
at the solar surface (negative in the north, positive in the south),
where significant amounts of magnetic helicity can be ejected via
coronal mass ejections.
It is shown that open boundaries tend to alleviate catastrophic alpha quenching.
The importance of looking at current helicity instead of magnetic helicity
is emphasized and the conceptual advantages are discussed.
\end{abstract}

\firstsection % if your document starts with a section,
\section{Introduction}

The emerging magnetic field of the sun frequently displays strong signs
of twist.
The systematic investigation of twist began with the early work of
\cite{See90} who analyzed the current helicity in active regions and
found a hemispheric dependence of its sign: negative in the north,
positive in the south.
This dependence has since been confirmed and the statistics improved.
The investigation of helicity is usually motivated by the interest
in a detailed description of the degree of complexity of the solar
magnetic field.
There has also been some interest in understanding the reasons for the
observed twist (or helicity).
In recent years, however, a very different question has emerged:
how is it possible that the solar dynamo works as it does?
Given that in simulations the value of the Spitzer resistivity still
affects the cycle period (\cite{BDS02}),
one would like to understand how this is avoided in a proper theory
of the solar cycle.

The significance of this question is often not very evident, but this is
mainly because in many simulations the values of the magnetic Reynolds
number are still not large enough.
It should be emphasized that the involvement of the microscopic
diffusivity in the description of macroscopic properties of a turbulent
flow is a highly unusual property of MHD that is not normally encountered
anywhere else in turbulence.

What is so special here is that the magnetic helicity is an almost
perfectly conserved quantity.
This can have serious implications for the operation of the large scale
dynamo effect in the nonlinear regime.
An example where this is true is the $\alpha$ effect in
mean field electrodynamics (\cite{Mof78}; \cite{KR80}).
As more and more large scale field is produced by the $\alpha$ effect,
and since the large scale field produced by the $\alpha$ effect
is helical (with a sign equal to that of $\alpha$), there must
be a simultaneous generation of magnetic helicity of the opposite
sign such that the total magnetic helicity is conserved.
(What actually matters for the $\alpha$ effect is the {\it current
helicity}, not the magnetic helicity, but the two are related.)

The quenching of $\alpha$ can also affect the quenching of the
turbulent magnetic diffusivity and this in turn the cycle period.
Even though the cycle period of 22 years is significantly longer
than the local turnover time of the turbulence, we do not necessarily
expect the cycle period to depend on the magnetic diffusion time.
As pointed out earlier (\cite{BDS02}),
the idea of the cycle period being dependent
on the magnetic diffusion time is not completely unrealistic:
The magnetic helicity constraint dictates that the change
$\Delta H_{\rm N}$ of the
magnetic helicity during the solar cycle, normalized by the magnetic
energy $M_{\rm N}$, where the subscript N refers to the northern hemisphere,
must not exceed the skin depth if the change in magnetic helicity is
brought about by purely resistive effects
(the same applies separately for the southern hemisphere).
Models of the solar dynamo suggest that the ratio
$\Delta H_{\rm N}/(2\mu_0 M_{\rm N})$ is about $70\km$.
In the sun the skin depth based on the cycle period varies between $10\km$
at the bottom of the convection zone and $300\km$ at the top.
Thus, unless the dynamo works only in a thin surface layer,
$\Delta H_{\rm N}$ is too large compared to the $10\km$ figure.
Therefore one might hope that open boundaries can alleviate this
constraint, although the effect does not need to be very strong.

The effect of open boundaries has already been investigated in the past
in a model without differential rotation (\cite{BD01}).
It was found that open boundaries do lead to the expected reduction of the
saturation time scale, but the amplitude of the final field strength was
also reduced dramatically.
In the following we report on recent simulations of
\cite{BS} (2004, hereafter referred to as BS),
where the effect of open boundaries has
been investigated in a model with solar-like differential rotation.
This was found to lead to an increase of the saturation field strength by
a factor of about 30.
Such an increase was associated with the effect of a current helicity flux.
In the following we discuss their model in more detail and present new
calculations of the resulting dynamo action.

\section{A cartesian model of the solar differential rotation}
 
\begin{figure}\centering
\includegraphics[width=.99\textwidth]{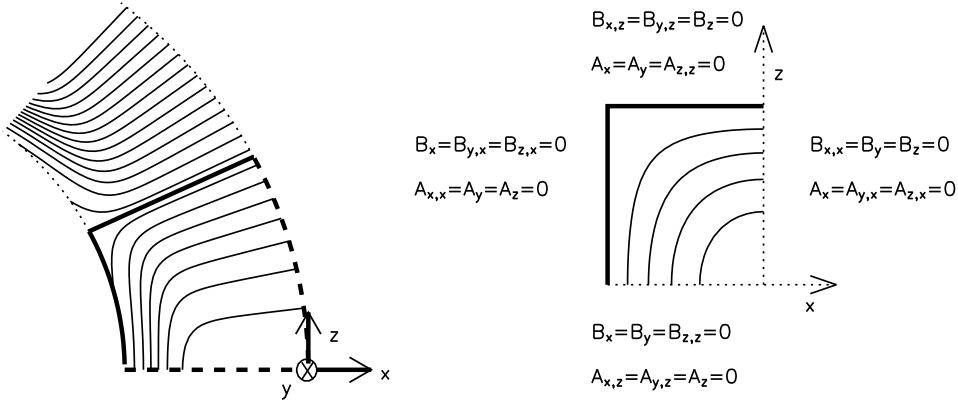}
\caption{
On the left hand side a
sketch of the solar angular velocity at low latitudes is shown, with
spoke-like contours in the bulk of the convection zone merging
gradually into uniform rotation in the radiative interior.
The low latitude region, considered in this paper, is
indicated by thick lines.
On the right hand side we show the
differential rotation as modeled in our cartesian box.
The equator corresponds to the bottom, the surface to the right,
the bottom of the convection zone to the left and mid-latitudes
at the top of the box.
The boundary conditions for the three components of the magnetic field
and the vector potential are indicated near the boundaries of the box.
}\label{sketch2}\end{figure}

In order to model the region below $30^\circ$ latitude,
BS have adopted a cartesian geometry where
the $x$ direction corresponds to radius, the $y$ direction to longitude,
and the $z$ direction to latitude.
The mean toroidal velocity is given by
\begin{equation}
\meanUU=U_0\cos k_1x\cos k_1z,
\label{shearprofile}
\end{equation}
where $k_1$ is the lowest wavenumber in the $(x,z)$ plane with
$-\pi/2\leq k_1x\leq0$ and $0\leq k_1z\leq\pi/2$.
In the following we adopt units where $k_1=1$.
The equator is assumed to be at $z=0$ and the outer surface at $x=0$.
The bottom of the convection zone is at $x=-\pi/2$ and the latitude where
the surface angular velocity equals the value in the radiative interior
is at $z=\pi/2$; see \Fig{sketch2}.

In order to test the properties of this differential rotation, BS
investigated first the resulting dynamo action in a mean field model.
This should give some idea of the type of solutions that one
might expect in a three-dimensional model where helical turbulence is able
to sustain dynamo action without explicitly invoking an $\alpha$ effect.

\begin{figure}\centering
\includegraphics[width=.9\textwidth]{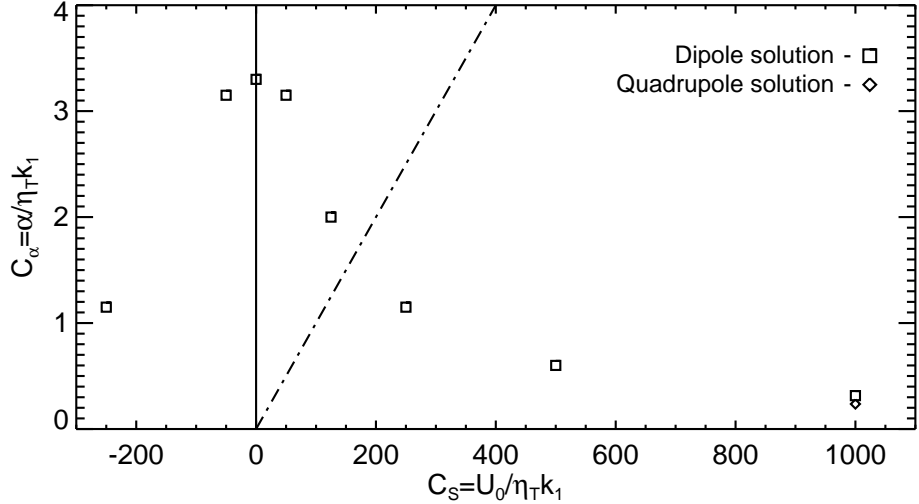}\caption{
Critical value of $C_\alpha$ for dynamo action as a function of
shear parameter, $C_S$.
Note the typical decrease of the critical value of $C_\alpha$ with
increasing $C_S$.
The dash-dotted line corresponds to the parameter regime
of the turbulence simulations discussed below.
}\label{growth}\end{figure}

In \Fig{growth} we plot the stability diagram in the $(C_\alpha,C_S)$ plane,
where $C_\alpha=\alpha/\eta_{\rm T}k_1$ and $C_S=U_0/\eta_{\rm T}k_1$ are
nondimensional measures of $\alpha$ effect and shear.
For $C_\alpha<C_{\alpha,{\rm crit}}$ the solutions are decaying and for
$C_\alpha>C_{\alpha,{\rm crit}}$ they are growing exponentially and are
oscillatory (Hopf bifurcation), except for a narrow interval around
$C_S=0$. 
Such a behavior is quite typical of $\alpha\Omega$ dynamos (see, e.g.,
\cite{RS72}). 
For $C_S=1000$ we have also considered the quadrupolar solution by
changing the boundary condition on the equator in the appropriate way.
It turns out that it is slightly easier to excite (see \Fig{growth}).
As stated earlier, the approximately equal excitation conditions for
dipolar and quadrupolar solutions, seen in \Fig{growth},
are typical of $\alpha\Omega$ dynamos in spherical shells.
Indeed, the fact that quadrupolar solutions can be preferred has been
found in other solar dynamo models (\cite{DG01}).

We may conclude that the present cartesian setup provides a useful
representation of a global model of the sun's differential rotation.
The mean field model reproduces similar features to those found
in global mean field models in spherical shells.

\section{Results for $\alpha$ quenching}
\label{ResultsAlphaQuenching}

\begin{figure}\centering
{\includegraphics[width=.9\textwidth]{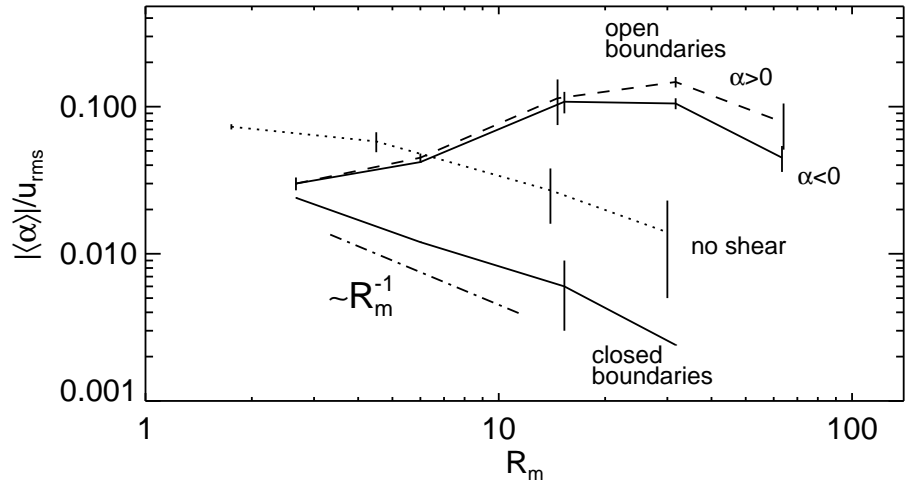}}\caption{
Dependence of $|\bra{\alpha}|/u_{\rm rms}$ on $R_{\rm m}$
for open and closed boundaries.
The case with open boundaries and negative helicity is shown as a dashed line.
Note that for $R_{\rm m}\approx30$ the $\alpha$ effect
is about 30 times smaller when the boundaries are closed.
The dotted line gives the result with open boundaries but no shear.
The vertical lines indicate the range obtained by calculating
$\alpha$ using only the first and second half of the time interval.
}\label{palp_sum}\end{figure}

Next, BS focussed on the investigation of helicity-driven turbulence in
the presence of shear as given by \Eq{shearprofile}.
Instead of looking for dynamo action, they considered the case of an
imposed magnetic field in the $y$ direction of strength $B_{0y}$.
They determined $\alpha$ by measuring the turbulent
electromotive force, i.e.\ $\alpha=\bra{\emf}\cdot\BB_0/B_0^2$.
They presented a range of simulations for different values of the
magnetic Reynolds number,
\EQ
R_{\rm m}=u_{\rm rms}/(\eta k_{\rm f}),
\EN
for both open and closed boundary conditions.
(Here, $u_{\rm rms}$ does not include the mean shear flow.)
In the simulations, $U_0/u_{\rm rms}\approx10$ and
$\alpha/u_{\rm rms}\approx0.1$ (\Fig{palp_sum}).
Using $\eta_{\rm T}=c_\eta u_{\rm rms}/k_{\rm f}$, where $c_\eta$ is
a free parameter, we have $C_S/C_\alpha\approx100$, which is marked in
\Fig{growth} as a dash-dotted line.
The intersection with the sequence of points from the mean field
calculation gives $C_S\approx150$, and since
\EQ
C_S\equiv{U_0\over\eta_{\rm T}k_1}
\approx{U_0 k_{\rm f}\over c_\eta u_{\rm rms}k_1}
\approx{50\over c_\eta}
\EN
and $k_{\rm f}/k_1=5$ we find $c_\eta\approx0.3$.
This value appears reasonable, although somewhat smaller that the
value of 0.8 obtained from magnetic decay experiments (\cite{YBR03}).
The value of $C_S\approx150$ suggests that our simulations should be in
an oscillatory regime, which is indeed confirmed (see \Sec{Dynamo}).

There is a striking difference between the cases with open and
closed boundaries which becomes particularly clear when comparing
the averaged values of $\alpha$ for different magnetic Reynolds
numbers; see \Fig{palp_sum}.
With closed boundaries $\alpha$ tends to zero like $R_{\rm m}^{-1}$,
while with open boundaries $\alpha$ shows no such immediate decline;
only for larger values of $R_{\rm m}$ there is possibly an asymptotic
$\alpha\propto R_{\rm m}^{-1}$ dependence.
There is also a clear difference between the cases with and without shear.
In the absence of shear (dotted line in \Fig{palp_sum}) $\alpha$ declines
with increasing $R_{\rm m}$, even though for small values of $R_{\rm m}$
it is larger than with shear.
This suggests that the presence of shear combined with open boundaries
might be a crucial prerequisite of dynamos that saturate on a dynamical
time scale.

The difference between open and closed boundaries
can be explained in terms of a current helicity
flux through the two open open boundaries of the domain.
Instead of going through the mathematical formalism, we just present the
argument in words.
First of all, in the kinematic regime (i.e.\ for weak fields) the
$\alpha$ effect is a negative multiple of the kinetic helicity.
As the magnetic field grows, there will also be a growing small scale
magnetic field which itself is helical and it too enters in the
calculation of $\alpha$.
The relevant quantity is the mean current helicity of the small scale,
$\overline{\jj\cdot\bb}$.
Its sign is that of the kinetic helicity, i.e.\ negative in the northern
hemisphere, and it enters with a minus sign, so it acts in such a way
as to quench the total $\alpha$ effect (\cite{PFL}).
The next important step is to find the evolution equation for
$\overline{\jj\cdot\bb}$ in terms of the mean field.
This can be done in the same way as in the calculation of the kinematic
$\alpha$ effect, e.g.\ by using the first order smoothing approach, or
by other techniques (see \cite[Brandenburg \& Subramanian 2004]{BS04}
for a review).
It is simpler, however, to use magnetic helicity conservation, so
one has to convert from current helicity, $\overline{\jj\cdot\bb}$,
to magnetic helicity, $\overline{\aaaa\cdot\bb}$.
Under isotropic conditions we have
$\overline{\jj\cdot\bb}=k_{\rm f}^2\,\overline{\aaaa\cdot\bb}$, where
$k_{\rm f}$ is the wavenumber of the fluctuating (small scale) field.
The time dependence of the magnetic helicity equation cannot usually
be ignored.
It is important to explain the slow saturation behavior found in helical
dynamos in closed and periodic boxes.
But after a resistive time scale (which can be very long) the time
dependence can be ignored.
In that case the magnetic helicity equation says that the production of
small scale magnetic helicity (which is equal and opposite in sign to
the production of large scale magnetic helicity and hence large scale
magnetic field) must be balanced by the magnetic helicity dissipation
term (i.e.\ the current helicity times the resistivity).
The latter term is hence resistively small, which is why the electromotive
force, and hence the $\alpha$ effect, are catastrophically quenched.
However, when there are open boundary conditions, the situation changes
and the electromotive force can now be balanced by the divergence of
the current helicity flux.
The fact that even for the open boundary conditions the curves tend to
bend downward might suggest that the current helicity flux term itself
could depend on the small scale magnetic diffusivity.
If this turns out to be the case, it may indicate that the vertical
field boundary conditions used here do not represent a sufficiently
realistic representation of the solar surface conditions.

\begin{figure}\centering
{\includegraphics[width=.8\textwidth]{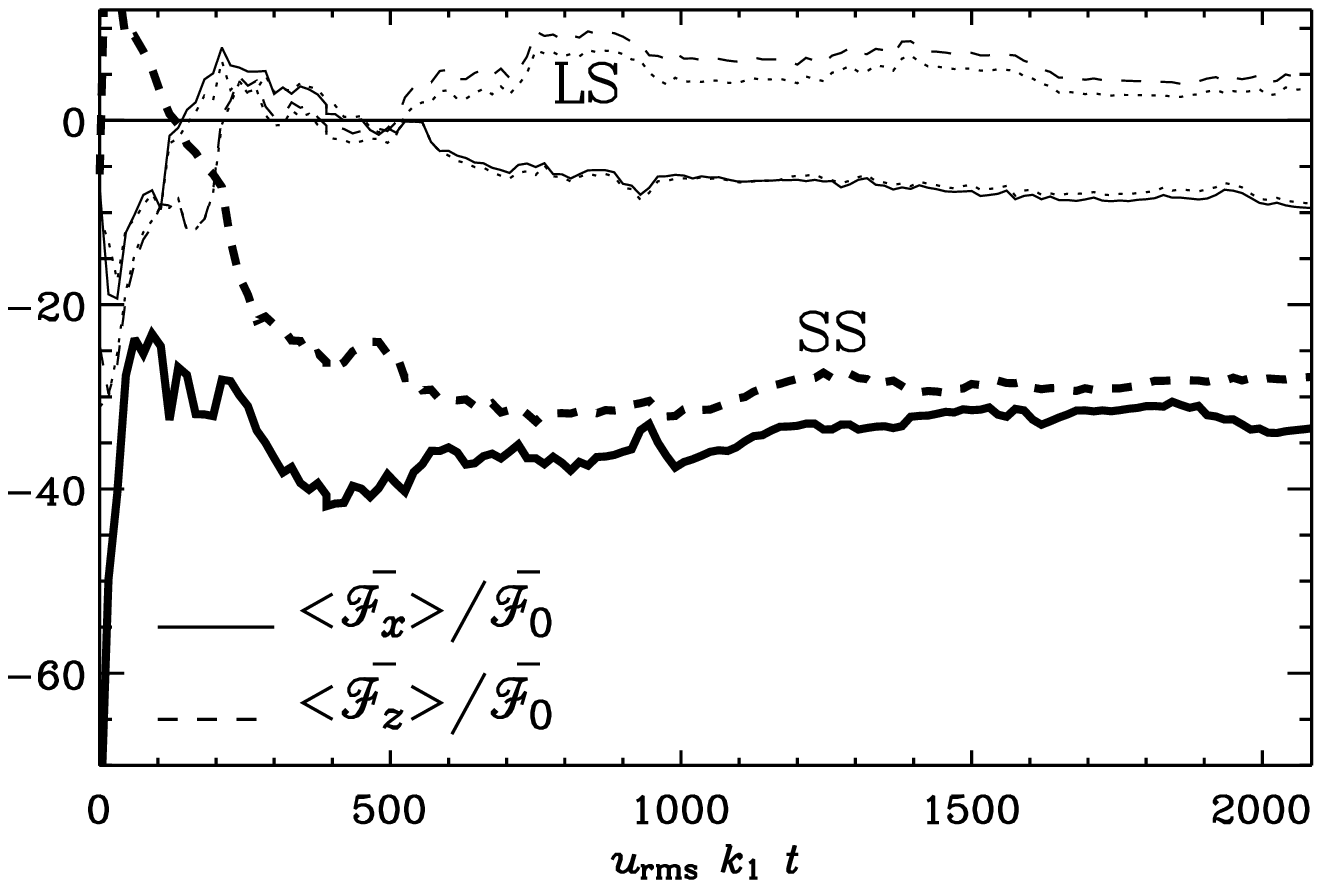}}\caption{
Normal components of the current helicity flux on the outer surface
($\meanF_x$) and at the equator ($\meanF_z$), averaged over the
corresponding surfaces.
The fat lines denote the fluxes from the small
scale field, $\meanFF_C^{\rm SS}$, while the thin lines denote
the fluxes form the large scale field, $\meanFF_C^{\rm LS}$.
The dotted lines near the two $\meanFF_C^{\rm LS}$ curves show
the result of the approximation
$\meanFF_C^{\rm LS}\approx-2(\meanJ_y\meanU_y)\meanBB$.
}\label{pcurhelflux}\end{figure} 

In \Fig{pcurhelflux} we also show the small scale
current helicity fluxes on the two boundaries (fat lines).
There is a tendency for the difference between incoming
flux at the equator (fat dotted line) and outgoing fluxes at outer surface
(fat solid line) to cancel, but the net outgoing flux is again negative.

A full investigation of the magnetic and current helicity losses
associated with dynamo action may really require global models in
spherical geometry.
However, some preliminary information can already now be gained by
studying local models with imposed solar-like differential rotation.
In mean field models such a geometry reproduces many of the features that
are known from corresponding mean field models in full spherical geometry.

Models with helically driven turbulence and an imposed toroidal magnetic
field allow the determination of the $\alpha$ effect.
It turns out that the catastrophic quenching of the $\alpha$ effect is
alleviated (at least by a factor of about 30) when magnetic and current
helicity is allowed to leave the domain.
The simulations have shown that a reasonable estimate
for the current helicity flux at the outer surface is
\EQ
\meanFF_C\approx30\,u_{\rm rms}k_{\rm f}B_0^2.
\EN
Applying this to the sun using $u_{\rm rms}\approx50\mpers$ for
the rms velocity in the deeper parts of the convection zone,
$k_{\rm f}\approx10^{-9}\cm^{-1}$ based on the inverse mixing length,
and $B_0\approx3\G$ for the mean field at the solar surface,
we have $\meanFF_C\approx10^{-3}\,\Gtwopers$. 
The current helicity flux integrated over the northern hemisphere of the sun
is then $4\times10^{19}\G^2\cm^2\s^{-1}$.
Integrated over the 11 yr solar cycle we have $10^{28}\G^2\cm^2$.

\begin{figure}\centering\includegraphics[width=0.4\textwidth]{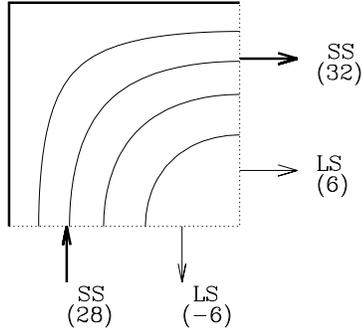}\caption{
Sketch illustrating the directions of large scale (LS) and small scale
(SS) negative current helicity fluxes and their approximate magnitudes
(in units of $\meanFF_0$).
Note that at the outer surface negative current helicity is ejected
both via small and large scale fields, while at the equator the
contributions from small and large scale fields have opposite sign.
The small scale losses at surface and equator partially cancel, giving
a net loss of negative current helicity of only about $4\,\meanFF_0$.
}\label{geom_flux}\end{figure}

For the sun only magnetic helicity fluxes have been determined.
As a rough estimate we may use
$\meanFF_H\approx k_{\rm f}^{-2}\meanFF_C$ for the
magnetic helicity flux. 
Using the same estimate for $k_{\rm f}$ as above we obtain
about $10^{46}\Mx^2$ over the 11 yr solar cycle.
This is indeed comparable to the magnetic helicity fluxes estimated
by \cite{BR00} and \cite{DV00}.

\section{Three-dimensional dynamo action}
\label{Dynamo}

\begin{figure}\centering\includegraphics[width=.9\textwidth]{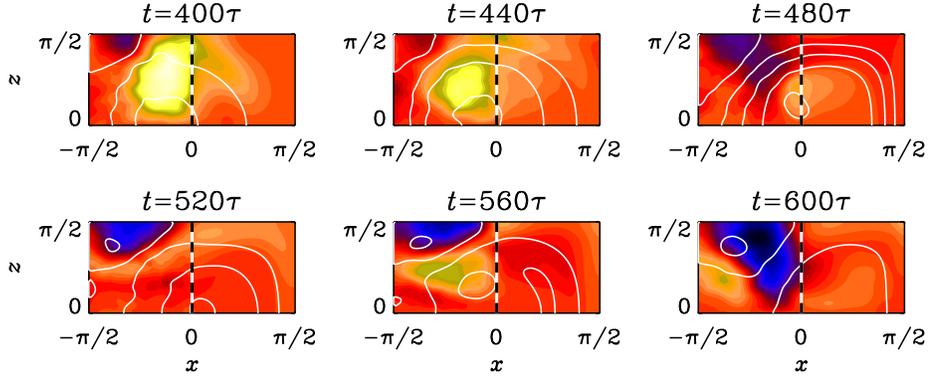}\caption{
The toroidally averaged field from a simulation without imposed field.
Poloidal field lines are superimposed on a color/grey scale representation
of the mean toroidal field.
The coronal buffer layer is to the right of the dashed line.
The frames are separated by about 40 turnover times, i.e.\
$\Delta t/(u_{\rm rms}k_{\rm f})=40$.
Note the migration of magnetic field toward the surface.
}\label{pslice64c}\end{figure}

In order to study the full dynamo operation we now present
simulations without imposed field.
In \Fig{pslice64c} we show an example of such a simulation where,
in addition to the helically driven turbulence (with negative helicity),
an outer coronal buffer layer has been added to allow the magnetic
field to be expelled from the turbulent dynamo zone.

\begin{figure}\centering\includegraphics[width=.9\textwidth]{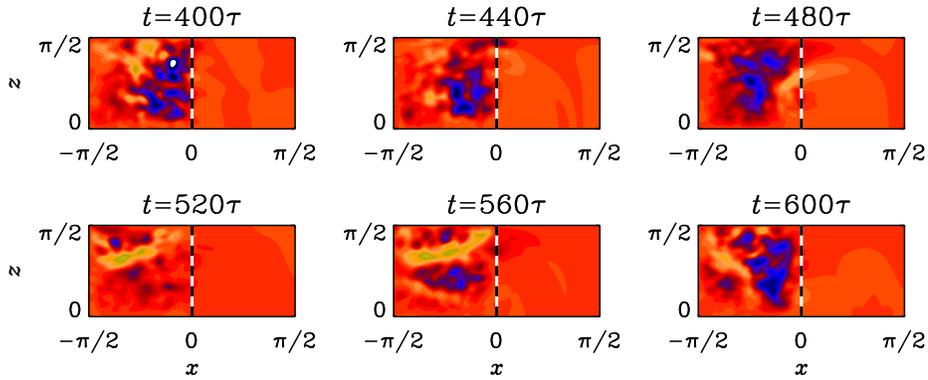}\caption{
Gray/color scale plot of the total current helicity density,
$\meanJJ\cdot\meanBB+\overline{\jj\cdot\bb}$ in the meridional plane.
Dark/blue represents negative values,
and intermediate/red shades indicate zero.
The coronal buffer layer is to the right of the dashed line.
Note that the total current helicity density is mostly negative,
but it's magnitude tends to be smaller near the boundaries.
}\label{pslice2_64c}\end{figure}

These studies are still preliminary and need to be carried out
for a range of different magnetic Reynolds numbers before we
are able to tell whether the open boundaries are really able
to alleviate the catastrophic $\alpha$ quenching that occurs
in the presence of closed or periodic boundaries; see Table~5
of \cite[Brandenburg et al.\ (2002)]{BDS02}.
Nevertheless, a few interesting aspects can already be recognized.
Firstly, there is some migration-like evolution of the toroidally
averaged magnetic field in the meridional plane -- similar to what
is expected from the mean field model (BS).
Secondly, by looking at the toroidally averaged current helicity
(mean and fluctuating parts together), one sees mostly negative
values (corresponding to dark shades in \Fig{pslice2_64c}).
On the open boundaries the current helicity is close to zero.
Only near the outer surface there is occasionally a dominance
of negative values.

\section{Conclusions}

Finally, in the absence of an imposed toroidal field a large
scale magnetic field is generated that shows features of
field migration toward the surface, similar to what the
mean field shows.
More work is obviously required to test the dependence of
the cycle period on the magnetic Reynolds number.
Also, it would be interesting to allow for a (nearly) force-free coronal
magnetic field that permits a more direct connection between the helicity
losses and the field geometry associated with these losses.
Ideally, of course, one would like to model proper coronal mass
ejections in the context of this model.

\begin{acknowledgments}
The Danish Center for Scientific Computing is acknowledged
for granting time on the Linux cluster in Odense (Horseshoe).
\end{acknowledgments}

\end{document}